\documentclass[final]{siamltex}

\usepackage{amssymb,amsmath, amsfonts,color,psfrag}
\usepackage[draft]{graphicx}
\usepackage[noend]{algorithmic}
\usepackage{subfigure}
\usepackage{epsfig}

\newcommand{\real}{{\mathbb{R}}}
\newcommand{\sphere}{{\mathbb{S}}}

\renewcommand{\natural}{{\mathbb{N}}}
\newcommand{\eps}{\epsilon}

\newcommand{\scirc}{\raise1pt\hbox{$\,\scriptstyle\circ\,$}}

\newcommand{\union}{\operatorname{\cup}}

\newcommand{\PP}{\mathcal{P}}

\newcommand{\MM}{\mathcal{M}}
\newcommand{\EE}{\mathcal{E}}

\newcommand{\s}{\mathcal{S}}

\newcommand{\VV}{\mathcal{V}} 

\newcommand {\Gvis}{\mathcal{G}_{\rm vis}}            
\newcommand{\Gcomm}{\mathcal{G}_{\rm comm}} 
\newcommand{\smax}{s_{\rm max}}                           
\newcommand{\interior}{\operatorname{int}}

\newcommand{\cball}[2]{\overline{B}_{#1}(#2)}

\newcommand{\LISTEN}{\textup{LISTEN}}
\newcommand{\SPEAK}{\textup{SPEAK}}
\newcommand{\PROCESS}{\textup{PROCESS}}

\newcommand{\SLEW}{\textup{SLEW}}
\newcommand{\BROADCAST}{\textup{BROADCAST}}
\newcommand{\RECEIVE}{\textup{RECEIVE}}

\newcommand\oprocendsymbol{\hbox{$\square$}}
\newcommand\oprocend{\relax\ifmmode\else\unskip\hfill\fi\oprocendsymbol}

\renewcommand{\theenumi}{(\roman{enumi})}

\begin{document}\renewcommand{\thefootnote}{\fnsymbol{footnote}}

\title{Asynchronous Distributed Searchlight Scheduling\footnotemark[1]}

\author{Karl Obermeyer\footnotemark[2], Anurag Ganguli\footnotemark[3] \and Francesco Bullo\footnotemark[2]}

\footnotetext[1]{Compiled \today.  Technical report for www.arXiv.org.  This work has been supported in part by AFOSR MURI Award F49620-02-1-0325, NSF Award CMS-0626457, and a DoD SMART fellowship. A preliminary version of this manuscript has been accepted for the 2007 Conference on Decision and Control, New Orleans, LA, USA.}
  
  \footnotetext[2]{Karl Obermeyer and Francesco Bullo are with the
    Department of Mechanical Engineering, University of California at Santa
    Barbara, Santa Barbara, CA 93106, USA, \texttt{karl@engr.ucsb.edu},
    \texttt{bullo@engineering.ucsb.edu}}
 
 \footnotetext[3]{Anurag Ganguli is with the Coordinated Science
   Laboratory, University of Illinois at Urbana-Champaign, and with the
   Department of Mechanical and Environmental Engineering, University of
   California at Santa Barbara, Santa Barbara, CA 93106, USA,
   \texttt{aganguli@uiuc.edu}}

\maketitle

\begin{abstract}
  This paper develops and compares two simple asynchronous distributed
  searchlight scheduling algorithms for multiple robotic agents in
  nonconvex polygonal environments.  A searchlight is a ray emitted by a
  agent which cannot penetrate the boundary of the environment.  A point is
  detected by a searchlight if and only if the point is on the ray at some
  instant.  Targets are points which can move continuously with unbounded
  speed.
  The objective of the proposed algorithms is for the agents to coordinate
  the slewing (rotation about a point) of their searchlights in a
  distributed manner, i.e., using only local sensing and limited
  communication, such that any target will necessarily be detected in
  finite time.  The first algorithm we develop, called the DOWSS
  (Distributed One Way Sweep Strategy), is a distributed version of a known
  algorithm described originally in 1990 by Sugihara et al
  \cite{KS-IS-MY:90}, but it can be very slow in clearing the entire
  environment because only one searchlight may slew at a time.  In an
  effort to reduce the time to clear the environment, we develop a second
  algorithm, called the PTSS (Parallel Tree Sweep Strategy), in which
  searchlights sweep in parallel if guards are placed according to an
  environment partition belonging to a class we call PTSS partitions.
  Finally, we discuss how DOWSS and PTSS could be combined with with deployment, or extended to environments with holes.
 \end{abstract}
\section{Introduction}
\label{sec:intro}
\begin{figure}[th]
\begin{center}
\resizebox{0.9\linewidth}{!}{ \includegraphics{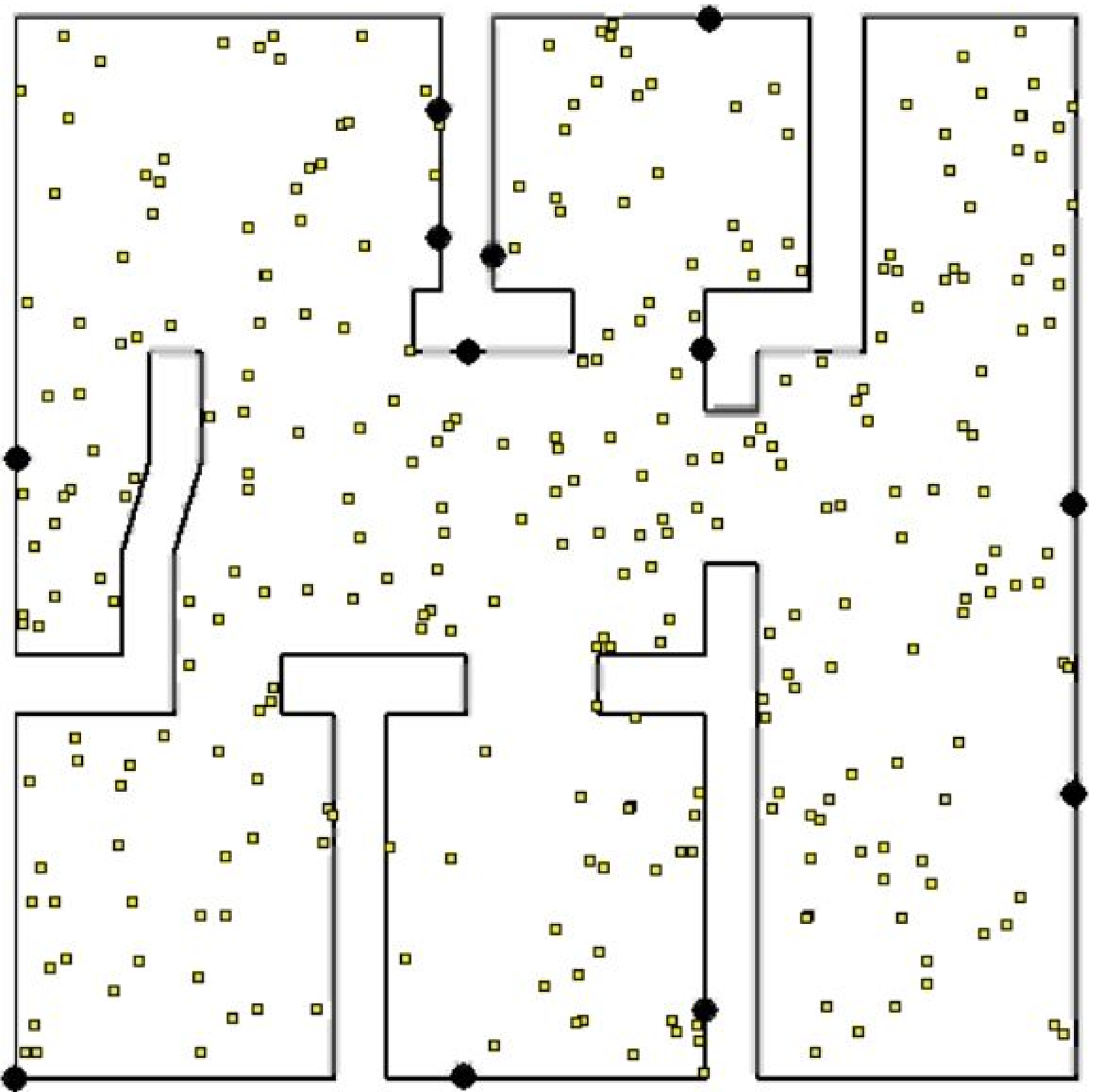} \qquad \includegraphics{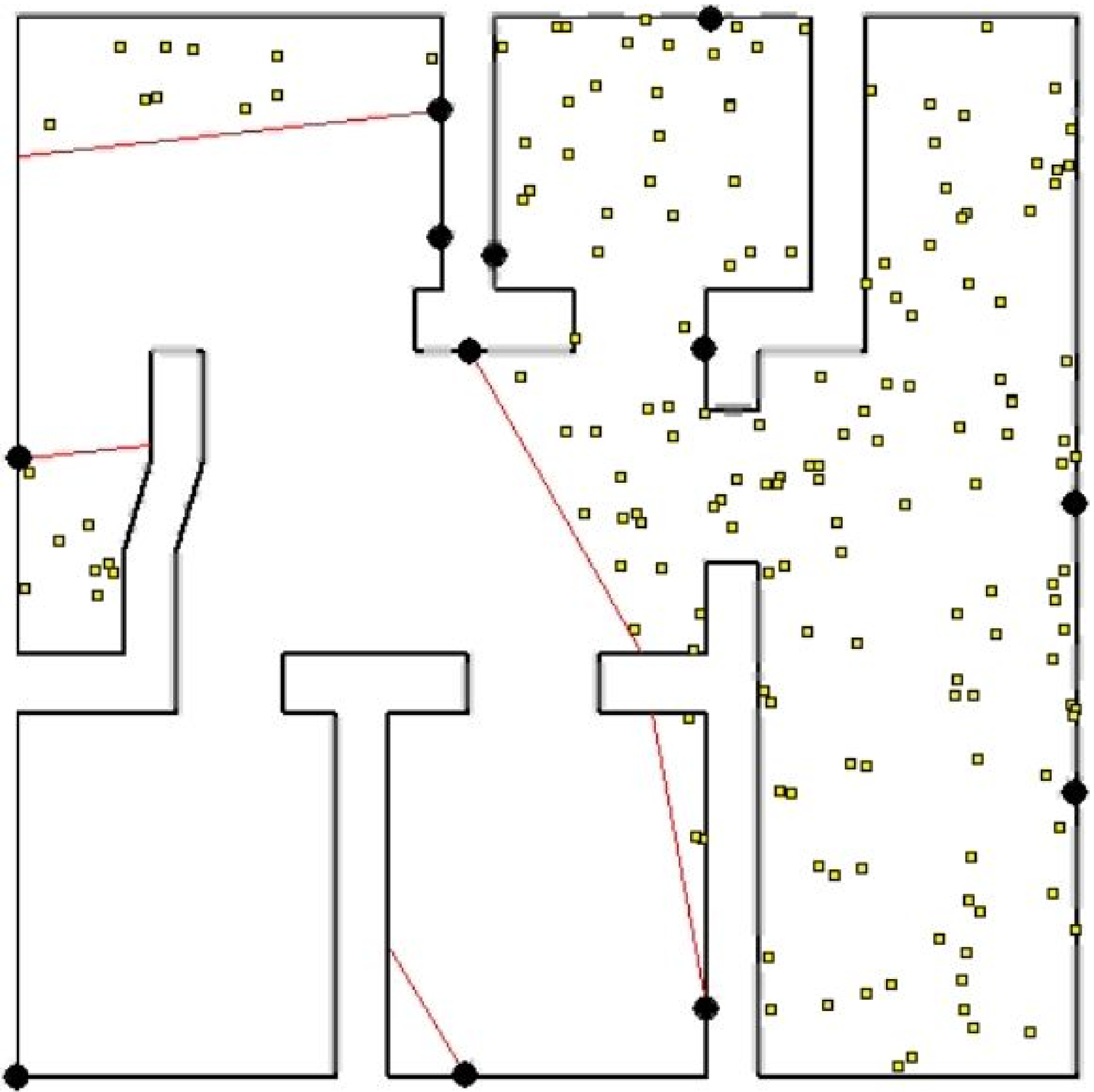} 
\qquad \includegraphics{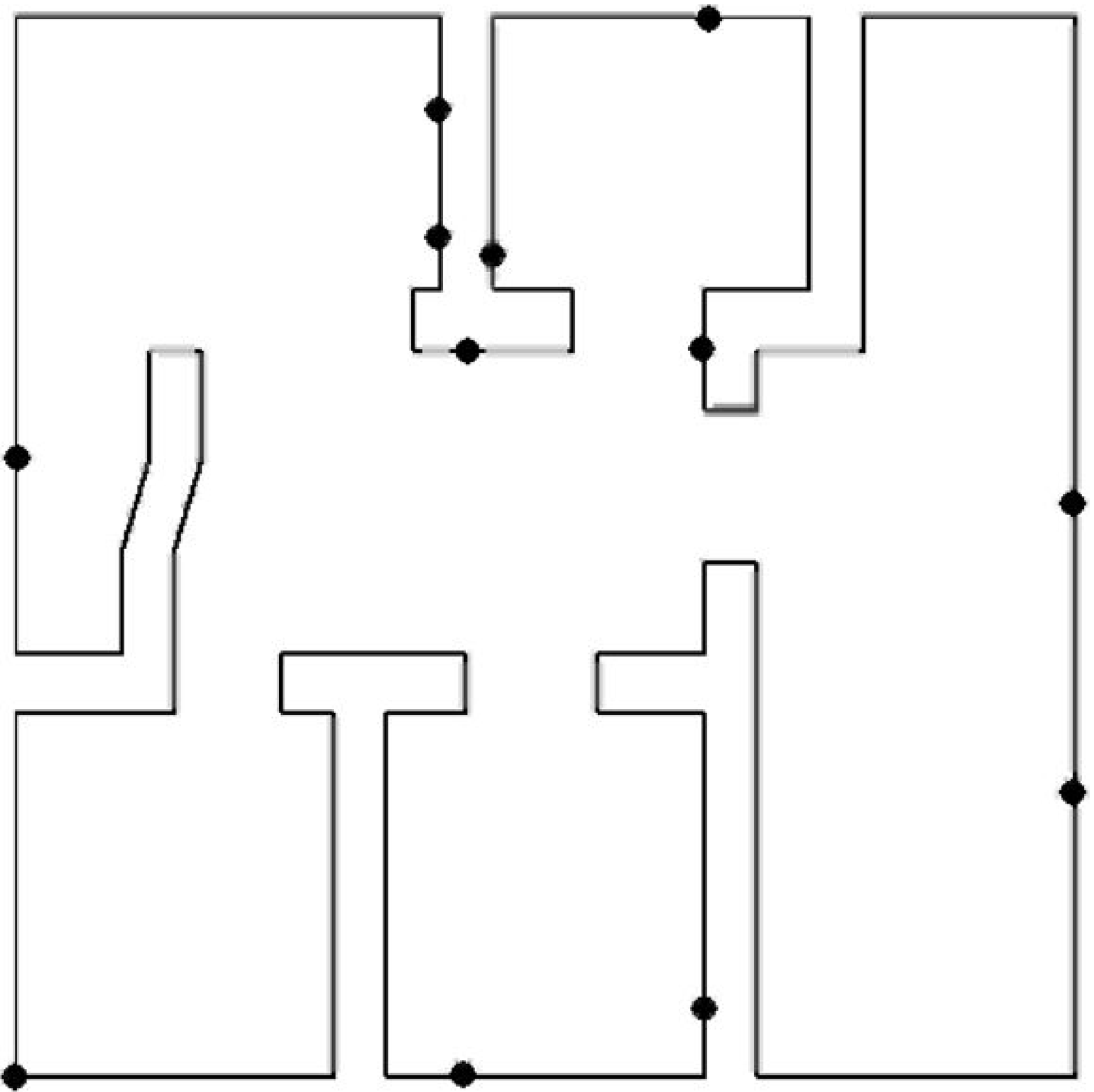} }




\caption{Simulation results of the PTSS algorithm described in Section~\ref{subsec:ptss}, executed by agents (black dots) in a polygon shaped like a typical floor plan.  Left to right 
, moving targets (small yellow squares) disappear as they are detected by searchlights (red).  The cleared region grows until it encompasses the entire environment.}
\label{fig:ptss_sim}
\end{center}
\end{figure}

\begin{figure}[h]
 \begin{center}
 \resizebox{0.5\linewidth}{!}{\includegraphics{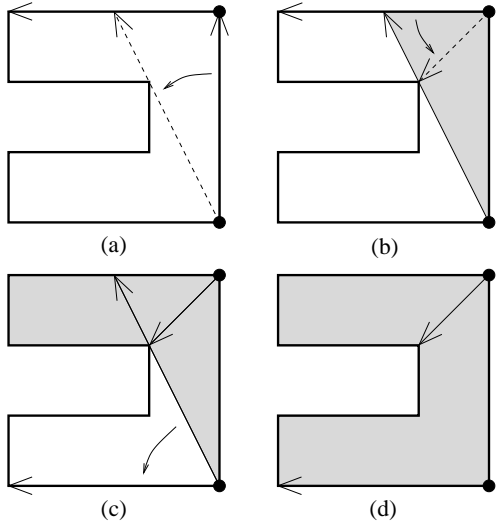}}
 \caption{A simple example of a searchlight schedule.  From (a) to (d):
   First the lower agent aims at the upper agent and sweeps until it hits a
   visibility gap.  Next, the upper agent sweeps the other side of the
   visibility gap so the lower agent can continue sweeping the remainder of
   the environment.  No target, no matter how fast, would be able to avoid
   being detected by this slewing sequence.}
 \label{fig:introexample}
 \end{center}
 \end{figure}

Consider a group of robotic agents acting as guards in a nonconvex
polygonal environment, e.g., a floor plan.  For simplicity, we model the
agents as point masses.  Each agent is equipped with a single
unidirectional sweeping sensor called a {\it searchlight} (imagine a ray of
light such as a laser range finder emanating from each agent).  A
searchlight aims only in one direction at a time and cannot penetrate the
boundary of the environment, but its direction can be changed continuously
by the agent.  A point is detected by a searchlight at some instant iff the
point lies on the ray.  A target is any point which can move continuously
with unbounded speed.  The \emph{Searchlight Scheduling Problem} is to
\begin{quote}
  Find a schedule to slew a set of stationary searchlights such that any
  target in an environment will necessarily be detected in finite time.
\end{quote} 
A searchlight problem instance consists of an environment and a set of
stationary guard positions.  Obviously there can only exist a search
schedule if all points in the environment are visible by some guard.  For a
graphical description of our objective, see Fig.~\ref{fig:ptss_sim} and
\ref{fig:introexample}.

To our knowledge the searchlight scheduling problem was first introduced in
the inspiring paper by Sugihara, Suzuki and Yamashita in
\cite{KS-IS-MY:90}, which considers simple polygonal environments and
stationary searchlights.  \cite{MY-IS-TK:04} extends \cite{KS-IS-MY:90} to
consider guards with multiple searchlights (they call a guard possessing $k$
searchlights a \emph{$k$-searcher}) and polygonal environments containing
holes.  Some papers involving mobile searchlights, sometimes calling them
\emph{flashlights} or \emph{beam detectors}, are \cite{BPG-ST-GG:06},
\cite{MY-HU-IS-TK:01}, \cite{BS-GS-SML:00}, 
and \cite{JHL-SMP-KYC:02}.  
  Closely related is the Classical Art Gallery Problem,
namely that of finding a minimum set of guards s.t. the entire polygon is
visible.  There are many variations on the art gallery problem which are
wonderfully surveyed in \cite{JU:00}, \cite{JOR:87}, and \cite{TCS:92}.

Assume now that each member of the group of guards is equipped with
omnidirectional line-of-sight sensors. By a line-of-sight sensor, we mean
any device or combination of devices that can be used to determine, in its
line-of-sight, (i) the position or state of another guard, and (ii) the
distance to the boundary of the environment.  By omnidirectional, we mean
that the field-of-vision for the sensor is $2\pi$ radians.  There exist
distributed algorithms to deploy asynchronous mobile robots with such
omnidirectional sensors into nonconvex environments, and they are
guaranteed to converge to fixed positions from which the entire environment
is visible, e.g., \cite{AG-JC-FB:05z} and \cite{AG-JC-FB:06r}.  At least one
algorithm exists which guarantees the ancillary benefit of the final guard
positions having a connected visibility graph (\cite{AG-JC-FB:06r}).

Once a set of guards seeing the entire environment has been established, it
may be desired to continuously sweep the environment with searchlights so
that any target will be detected in finite time.  The main contribution of
this paper is the development of two different asynchronous distributed
algorithms to solve the searchlight scheduling problem.  Correctness and 
bounds on time to clear nonconvex polygonal environments are discussed.  
The first algorithm we develop, called the DOWSS (Distributed
One Way Sweep Strategy, Sec.~\ref{subsec:dowss}, is a distributed version
of a known algorithm described originally in \cite{KS-IS-MY:90}, but it can
be very slow in clearing the entire environment because only one
searchlight may slew at a time.  On-line processing time required by agents
during execution of DOWSS is relatively low, so that the expedience with
which an environment can be cleared is essentially limited by the maximum
angular speed searchlights may be slewn at.  In an effort to reduce the
time to clear the environment, we develop a second algorithm, called the
PTSS (Parallel Tree Sweep Strategy, Sec.~\ref{subsec:ptss}), which sweeps
searchlights in parallel if guards are placed according to an environment
partition belonging to a class we call PTSS partitions.  That we analyze
the time it takes to clear an environment, given a bound on the angular
slewing velocity, is a unique feature among all papers involving
searchlights to date.  Finally, we discuss how DOWSS and PTSS can be
extended for environments with holes and for mobile guards performing a
coordinated search.  Until now, there has been no description in the
literature of a scalable distributed algorithm for clearing an environment
with mobile searchlights (1-searchers), though \cite{BPG-ST-GG:06} and
\cite{MY-HU-IS-TK:01}, for example, offer some centralized approaches.

We begin with some technical definitions, statement of assumptions, and
brief description of the known centralized algorithm called the one way
sweep strategy (appears, e.g., in \cite{KS-IS-MY:90}, \cite{MY-IS-TK:04},
\cite{MY-HU-IS-TK:01}).  We then develop a partially asynchronous model, a
distributed one way sweep strategy, and our new algorithm the parallel tree
sweep strategy.

\section{Preliminaries}
\label{sec:prob_setup}
\subsection{Notation}
We begin by introducing some basic notation. We let $\real$, $\sphere^1$, 
and $\natural$ refer to the set of real numbers, the circle, and natural numbers, respectively.  
Given two points $x,y \in \real^2$, we let $[x,y]$ signify the \emph{closed 
segment} between $x$ and $y$.  Similarly, $]x,y[$ is the \emph{open
  segment} between $x$ and $y$, $[x,y[$ represents the set $]x,y[ \union
\{x\}$ and $]x,y]$ is the set $]x,y[ \union \{y\}$. Given a finite set $X$,
let $|X|$ represent the cardinality of the set. Also, we shall use $P$ to
refer to tuples of elements in $\real^2$ of the form $(p^{[0]},\dots,
p^{[N-1]})$ (these will be the locations of the agents), where $N$ denotes
the total number of agents.

We now turn our attention to the environment we are interested in and to
the concepts of visibility in such environments. Let $\EE$ be a simple
polygonal environment, possibly nonconvex. By simple, we mean that $\EE$ does
not contain any hole and the boundary does not intersect itself.
Throughout this paper, $n$ will refer to the number of edges of $\EE$ and $r$
the number of reflex vertices.  A point $q\in \EE$ is \emph{visible from}
$p\in \EE$ if $[p,q] \subset \EE$.  The \emph{visibility set} $\VV(p)\subset \EE$
from a point $p \in \EE$ is the set of points in $\EE$ visible from $p$.  A
\emph{visibility gap} of a point $p$ with respect to some region $R \subset
\EE$ is defined as any line segment $[a,b]$ such that $]a,b[ \subset
\interior(R)$, $[a,b] \subset \partial \VV(p)$, and it is maximal in the
sense that $a,b \in \partial R$ (intuitively, visibility gaps block off
portions of $R$ not visible from $p$).  The visibility graph $\Gvis$ of a
set of agents $P$ in environment $\EE$ is the undirected graph with $P$ as
the set of vertices and an edge between two agents iff they are visible to
each other.

We now introduce some notation specific to the searchlight problem. An
instance of the searchlight problem is specified by a pair $(\EE,P)$,
where $\EE$ is an environment and $P$ is a set of searchlight
locations in $\EE$.  For convenience, we will refer to
the $i$th searchlight as $s^{[i]}$ (which is located at $p^{[i]} \in
\real^2$), and $S=\{s^{[0]},\dots,s^{[N-1]}\}$ will be the set of all
searchlights.  $\theta^{[i]}$ will denote the configuration angle of
the searchlight in radians from the positive horizontal axis, and
$\Theta=\{\theta^{[0]},\dots,\theta^{[N-1]}\}$ the joint
configuration.  So, if we say, e.g., aim $s^{[i]}$ at point $e$, what
we really mean is set $\theta^{[i]}$ equal to an angle such that the
$i$th searchlight is aimed at $e$.  Note that searchlights do not
block visibility of other searchlights.

The next few definitions were taken from \cite{KS-IS-MY:90}.
\begin{definition}[schedule]
\label{defn:schedule}
The \emph{schedule} of a searchlight $s^{[i]} \in S$ is a continuous function
$\theta^{[i]}:[0,t^*] \mapsto \sphere^1$, where $[0,t^*]$ is an interval of real time.
\end{definition}

The \emph{ray} of $s^{[i]}$ at time $t \in [0,t^*]$ is the intersection of $\VV
(p^{[i]})$ and the semi-infinite ray starting at $p^{[i]}$ with direction $\theta^{[i]}(t)$.
$s^{[i]}$ is said to be \emph{aimed} at a point $x \in \EE$ in some time instant
if $x$ is on the ray of $s^{[i]}$.  A point $x$ is \emph{illuminated} if there
exists a searchlight aimed at $x$.

\begin{definition}[separability]
\label{defn:separability}
Two points in $\EE$ are \emph{separable} at time $t \in [0,t^*]$ if every
curve connecting them in the interior of $\EE$ contains an illuminated point,
otherwise they are \emph{nonseparable}.
\end{definition}

\begin{definition}[contamination and clarity]
\label{defn:contamination_clarity}
A point $x \in \EE$ is \emph{contaminated} at time zero if and only if it is
not illuminated. The point $x$ is contaminated at time $t \in ]0,t^*]$ iff
$\exists y \in \EE$ such that (1) $y$ is contaminated at some $t' \in [0,t[$,
(2) $y$ is not illuminated at any time in the interval $[t',t]$, and (3)
$x$ and $y$ are nonseparable at $t$.  A point which is not contaminated is
called \emph{clear}.  A region is said to be \emph{contaminated} if it
contains a contaminated point, otherwise it is \emph{clear}.
\end{definition}  



\begin{definition}[search schedule]
\label{defn:search_schedule}
Given $\EE$ and a set of searchlight locations $P = \{p^{[0]},\ldots, p^{[N-1]}\}$, the set $\Theta=\{\theta^{[0]}, \ldots, \theta^{[N-1]}\}$ is a \emph{search schedule for $(\EE,P)$} if
$\EE$ is clear at $t^*$.
\end{definition}

\subsection{Problem description and assumptions}
We now describe the problem we solve and the assumptions made.  The \emph{Distributed Searchlight Scheduling Problem} is to
\begin{quote}
  Design a distributed algorithm for a network of autonomous robotic agents
  in fixed positions, who will coordinate the slewing of their searchlights so that any target in an environment will necessarily be
  detected in finite time.  Furthermore, these agents are to operate using
  only information from local sensing and limited communication.
\end{quote}
What is precisely meant by local sensing and limited communication will
become clear in later sections.  The following \emph{standing assumptions}
will be made about every searchlight instance in this paper:
\begin{enumerate}
\vspace{0.5em}
\item The environment is a simple polygon with finitely many reflex vertices.
\\ {\it Comments:}  Compactness is a practical assumption for sensor range limitations.   Simple connectedness means no holes.  Having only finitely many reflex vertices precludes problems such as arise from fractal environments and will be important for proving the algorithms terminate in finite time.  
\vspace{0.5em}
\item Every point in the environment is visible from some agent and there
  are a finite number $N \in \natural$ of agents.  
  \\ {\it Comments:} If there were some point in the environment not visible by any agent, then a target could remain there undetected for infinite time.  
  \vspace{0.5em}
\item For every connected component of $\Gvis$, there is at least one agent
  located on the boundary of the environment.  
\\ {\it Comments:} This will be important for proving the algorithms terminate without failure.  It also implies every agent is either on the boundary of the environment or visible from some other agent.  If there existed an agent $i$ located at a point $p_i$ in the interior of the environment and not visible by any other agent, then there would exist $\eps >0$ such that $\cball{\eps}{p_i} \cap \VV(p_j) = \emptyset$ for $i \neq j$.  A target could thus evade detection by remaining in $\cball{\eps}{p_i}$ and simply staying on the opposite side of agent $i$ as $l_i$ points.
\end{enumerate}

\subsection{One Way Sweep Strategy (OWSS)}
\label{subsec:owss}
This section describes informally the centralized recursive One Way Sweep
Strategy (OWSS hereinafter) originally introduced in
\cite{KS-IS-MY:90}. The reader is referred to \cite{KS-IS-MY:90} for a detailed description.  Centralized OWSS also appears in
\cite{MY-HU-IS-TK:01} and \cite{MY-IS-TK:04}.  OWSS is a method for
clearing a subregion of a simple 2D region $\EE$ determined by the rays of
searchlights.  The subregions of interest are the so-called
\emph{semiconvex subregions} of $\EE$ \emph{supported} by a set of
searchlights at a given time and are defined as follows:

\begin{definition}[semiconvex subregion]
\label{defn:semiconvex_subreg} 
$\EE$ is always a semiconvex subregion of $\EE$ supported by $\emptyset$.  Furthermore, any $R \subset \EE$ is a semiconvex subregion of $\EE$ supported by a set of searchlights $S_{\rm sup}$ if both of the following hold:
\begin{enumerate}
\item It is enclosed by a segment of $\partial \EE$ and the rays of some of the searchlights in $S_{\rm sup}$.
\item The interior of $R$ is not visible from any searchlight in $S_{\rm sup}$.
\end{enumerate}
\end{definition}

\begin{figure}[h]
 \begin{center}
 \resizebox{0.7\linewidth}{!}{\input{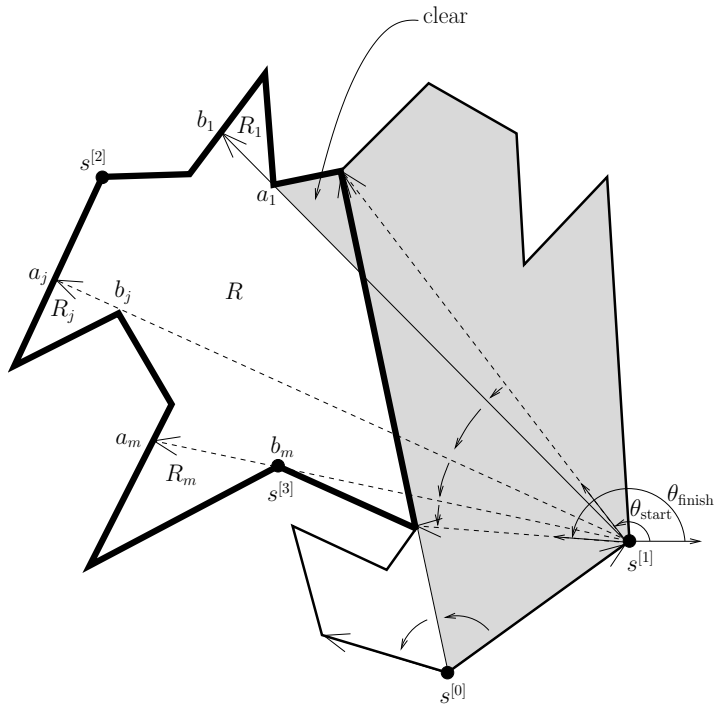}}
 \caption{One Way Sweep Strategy (OWSS) 
clears, by slewing $s^{[1]}$, the semiconvex subregion $R$ (thick border)
supported by $s^{[0]}$.  $\ s^{[1]}$ must stop incrementally at each of its
visibility gaps $[a_1, b_1]$, $[a_j, b_j]$, and $[a_m, b_m]$.  In this
recursive process, the regions ($R_1$, $R_j$, $R_m$) behind the visibility
gaps become semiconvex subregions supported by $\{s^{[0]}, s^{[1]}\}$, and must be
cleared using only the remaining searchlights ($s^{[2]}$ and $s^{[3]}$)
.  
}
 \label{fig:owss}
 \end{center}
 \end{figure}

The term ``semiconvex" comes from the fact that any reflex vertex of a
semiconvex subregion is also a reflex vertex of $\EE$.  In polygonal
environments, all semiconvex subregions are polygons.  The schedule used in
Fig. \ref{fig:introexample} was based on OWSS, but as a more general
example, consider Fig. \ref{fig:owss}. To clear the environment $\EE$, which
is a semiconvex subregion supported by $\emptyset$, we may begin by
selecting an arbitrary searchlight on the boundary, say $s^{[0]}$.  The
first searchlight selected to clear an environment will be called the
\emph{root}.  $s^{[0]}$ aims as far clockwise (cw hereinafter)
as possible so that it is aligned
along the cw-most edge.  $l_i$ will then slew couterclockwise (ccw hereinafter) through the environment, stopping incrementally whenever it encounters a visibility gap.
The only visibility gap $s^{[0]}$ encounters produces the semiconvex subregion
$R$ (thick border).  At this time, another searchlight which sees across
the visibility gap and is not in the interior of $R$, in this case $s^{[1]}$,
is chosen to begin sweeping the area in $R$ not seen by $s^{[0]}$.  Notice we
have marked angles $\phi_{\rm start}$ and $\phi_{\rm finish}$.  These are
the cw-most and ccw-most directions, resp., in which $s^{[1]}$ can aim at some
point in $R$.  $s^{[1]}$ will slew from $\phi_{\rm start}$ to $\phi_{\rm
  finish}$ and in the process encounter visibility gaps, each producing the
semiconvex subregions $R_1$, $R_j$, and $R_m$, which must be cleared by
$s^{[2]}$ and/or $s^{[3]}$.  As soon as $R$ is clear (when $\theta^{[1]}=\phi_{\rm
  finish}$), $s^{[0]}$ can continue slewing until it is pointing along the wall
immediately to its left at which time the entire environment is clear.  The
recursive nature of OWSS should be apparent at this point.  Note that
in OWSS (and DOWSS described later) it is actually arbitrary whether a
searchlight slews cw or ccw over a semiconvex subregion, but to simplify
the discussion we always use ccw.

\section{Asynchronous Network of Agents with Searchlights} 
\label{sec:asynch_model}

In this section we lay down the sensing and communication framework for the searchlight equipped agents which will be able to execute the proposed algorithms.  Each agent is 
able to sense the relative position of any point in its visibility set as well as identify visibility gaps on the boundary of its visibility set.  The agents' communication graph $\Gcomm$ is assumed to connected.  An agent can slew its searchlight  continuously in any direction and turn it on or off.

Each of the $N$ agents has a unique identifier (UID), say $i$, and a portion of memory dedicated to outgoing messages with contents denoted by $\MM^{[i]}$.  Agent $i$ can broadcast its UID together with $\MM^{[i]}$ to all agents within its communication region, where the communication region is defined differently in each algorithm.  Such a broadcast will be denoted by $\BROADCAST(i, \MM^{[i]})$.  
We assume a bounded time delay, $\delta>0$, between a broadcast and the corresponding reception.

Each agent repeatedly performs the following sequence of actions between any two wake-up instants, say instants $T^{[i]}_l$ and $T^{[i]}_{l+1}$ for agent $i$:
\begin{enumerate}
\item $\SPEAK$, that is, send a $\BROADCAST$ repeatedly at $\delta$ intervals, until it starts slewing;
\item $\LISTEN$ for a time interval at least $\delta$;
\item $\PROCESS$ and $\LISTEN$ after receiving a valid message;
\item $\SLEW$ to an angle decided during $\PROCESS$.
\end{enumerate}

\begin{figure}[thpb]
  \begin{center}
    \resizebox{0.75\linewidth}{!}{\input{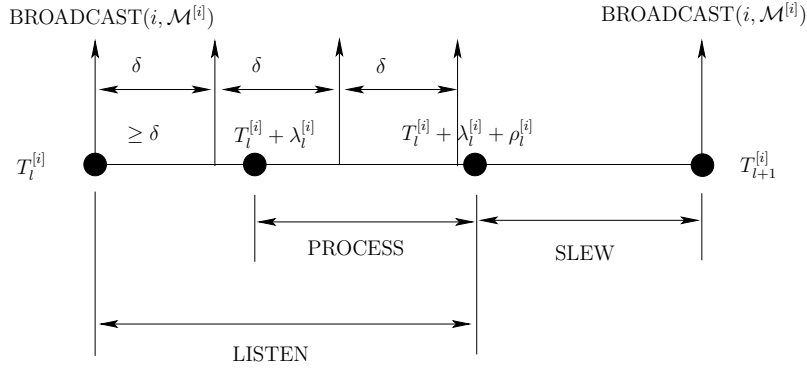}}
  \end{center}
  \caption{Sequence of actions performed by an agent $i$ in between two
    wake-up instants. Note that a $\BROADCAST(i,\MM^{[i]})$ is an instantaneous
    event taking place where there is a vertical pulse, where as the
    $\PROCESS$, $\LISTEN$ and $\SLEW$ actions take place over an interval.
    The $\SLEW$ interval may be empty if the agent does not sweep.}
  \label{fig:procschedule}
\end{figure}

See Figure~\ref{fig:procschedule} for a schematic illustration of the above schedule.

Any agent $i$ performing the $\SLEW$ action does so according to the following
discrete-time control system (cf Section~\ref{subsec:owss}):
\begin{equation}\label{eq:agent-dynamics}
  \theta^{[i]}(t+\Delta t) = \theta^{[i]}(t) + u^{[i]},
\end{equation}
where the control is bounded in magnitude by $\smax$.  The control action depends on time, values of variables stored in local memory, and the information obtained from communication and sensing.  The subsequent wake-up instant $T_{l+1}^{[i]}$ is the time when the agent stops performing $\SLEW$ and is not predetermined.  This network model is identical to that used for distributed deployment in \cite{AG-JC-FB:05z} and \cite{AG-JC-FB:06r}, and is similar in spirit to the \emph{partially asynchronous model} described in~\cite{DPB-JNT:97}.

\section{Distributed Algorithms}
\label{sec:algorithms}
In this section we design distributed algorithms for a network of agents
as described in Section~\ref{sec:asynch_model}, where no agent has global
knowledge of the environment or locations of all other agents. 

\subsection{Distributed One Way Sweep Strategy (DOWSS)}
\label{subsec:dowss}
Once one understands OWSS as in Section~\ref{subsec:owss}, esp.
its recursive nature, performing one way sweep of an environment in a
distributed fashion is fairly straightforward.  We give here an
informal description and supply a pseudocode in
Tab.~\ref{tab:dowss_short} (A more detailed pseudocode, which we refer to in the proofs, can be found in the appendix, Tab.~\ref{tab:dowss}).  In our discussion root/parent/child will refer to the relative location of agents in the simulated one way sweep recursion
tree.  In this tree, each node corresponds to a one way slewing action by
some agent.  A single agent may correspond to more than one node, but only
one node at a time
.  To begin DOWSS, some agent (the root\footnote{
The root could be chosen by any leader election scheme, e.g., a 
predetermined or lowest UID.}), say $i$, can aim as far cw as
possible and then begin slewing until it encounters a visibility gap.
Paused at a visibility gap, agent $i$ broadcasts a call for help to the
network.  For convenience, call the semiconvex subregion which $i$ needs
help clearing $R$.  All agents not busy in the set of supporting
searchlights $S_{\rm sup}$ (indeed at the zeroth level of recursion only the root is
in $S_{\rm sup}$), who also know they can see a portion of $\interior(R)$ but are not
in $\interior(R)$, volunteer themselves to help $i$.  Agent $i$ then
chooses a child and the process continues recursively.  In DOWSS as in 
Tab.~\ref{tab:dowss_short}, an agent needing help always chooses the
first child to volunteer, but some other criteria could be used, e.g.,
who sees the largest portion of $R$.  Whenever a child is finished
helping, i.e., clearing a semiconvex subregion, it reports to its parent so
the parent knows they may continue slewing.

\begin{figure}[h]
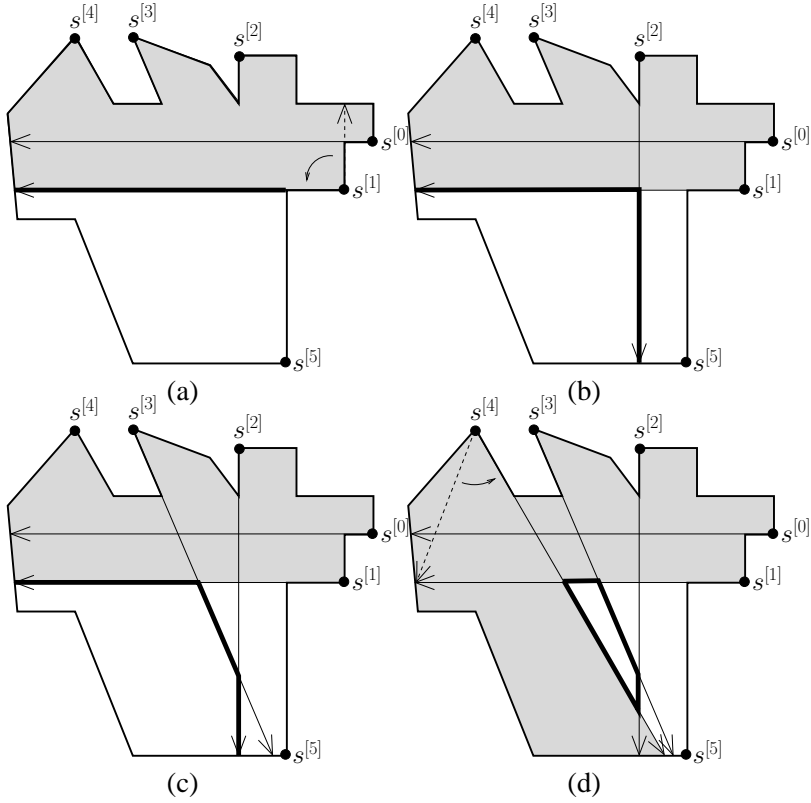

\begin{center}
\resizebox{0.8\linewidth}{!}{ \input{./fig/dowss2.tex} \qquad \input{./fig/dowss3.tex} }
\resizebox{0.8\linewidth}{!}{ \input{./fig/dowss4.tex} \qquad \input{./fig/dowss5.tex} }
\caption{An example execution of DOWSS.  The configuration in (a) results
  from $s^{[0]}$ clearing the very top of the region with help of $s^{[2]}$, $s^{[3]}$,
  and $s^{[4]}$ followed by $s^{[1]}$ attempting to clear the semiconvex subregion
  below where $s^{[0]}$ is aimed.  When $s^{[1]}$ gets stuck, it requests help by
  broadcasting the thick black polyline in (a), in this case just a line
  segment.  $s^{[2]}$ then helps $s^{[1]}$ but gets stuck right off, so it
  broadcasts the thick black polyline shown in (b).  Next $s^{[3]}$ helps $s^{[2]}$
  but gets stuck and broadcast the polyline in (c).  Similarly $s^{[4]}$
  broadcasts the polyline in (d), in this case a convex polygon, which only
  $s^{[5]}$ can clear.  In general, information passed between agents during
  any execution of DOWSS will be in the form of either an oriented line
  segment (a), a general oriented polyline (b and c), or  a convex polygon
  (d). }
\label{fig:dowss}
\end{center}
\end{figure}

The only subtle part of DOWSS is getting agents to recognize, without
global knowledge of the environment, that they see the interior of a
particular semiconvex subregion which some potential parent needs help
clearing.  More precisely, suppose some agent $l$ must decide whether to
respond as a volunteer to agent $i$'s help request to clear a semiconvex
subregion $R$.  Agent $l$ must calculate if it actually satisfies the
criterion in Tab.~\ref{tab:dowss}, line 3 of PROCESS, namely $p^{[l]}
\notin \interior(R)$ and $\interior(R) \cap \VV(p^{[l]}) \neq \emptyset$.
This is accomplished by agent $i$ sending along with its help request an
oriented polyline $\psi$ (see Tab.~\ref{tab:dowss}, line 3 of SPEAK).  By
an oriented polyline we mean that $\psi$ consists of a set of points listed according to some orientation convention, e.g., so that if one were to walk along the points in the order listed, then the interior of $R$ would always be to the right.  The polyline encodes the portion of $\partial
R$ which is not part of $\partial \EE$ and the orientation encodes which side
of $\psi$ is the interior of $R$.  Notice that for this to work, all agents
must have a common reference frame. 
Whenever the root broadcasts a polyline, it is just a line segment, but as
recursion becomes deeper, an agent needing help may have to calculate a
polyline consisting of a portion of its own beam and its parent's polyline.  The polyline may even close on
itself and create a convex polygon.  Examples of these scenarios are illustrated by in Fig~\ref{fig:dowss}.  We conclude our description of DOWSS
with the following theorem.
\begin{theorem}[Correctness of DOWSS]
\label{thm:dowss_correctness}
Given a simple polygonal environment $\EE$ and agent positions
$P=(p^{[0]},\ldots, p^{[N-1]})$, let the following conditions hold:
\begin{enumerate}
\item the standing assumptions are satisfied;
\item all agents $i \in \{0,\ldots,N-1\}$ have a common reference frame;
\item $p^{[0]} \in \partial \EE$;
\item the agents operate under DOWSS. 
\end{enumerate}
Then $\EE$ is cleared in finite time.
\end{theorem}
 \begin{proof}
   As in Theorem 2 of \cite{KS-IS-MY:90}, whenever an agent, say $i$, needs
   help clearing a semiconvex subregion $R$, there is some available agent
   $l$ satisfying $p{[l]} \notin \interior(R)$ and $\interior(R) \cap \VV(p^{[l]})
   \neq \emptyset$.  This comes from the standing assumption that for every
   connected component of $\Gvis$, there is at least one agent on the
   environment boundary.  Now since visibility sets are closed, we may
   demand additionally that agent $l$ sees a portion of the oriented
   polyline $\psi$ sent to it by $i$.  This means that in an execution of
   DOWSS, some $l$ will always be able to recognize, using only knowledge of
   $\VV(p^{[l]})$ and $\psi$ from local sensing and limited communication, that
   it is able to help.  We conclude DOWSS simulates OWSS.
 \end{proof}
 
We now give an upper bound on the time it takes DOWSS to clear the environment
assuming the searchlights slew at some constant angular velocity $\omega$, and that communication and processing time are negligible.  

 \begin{lemma}[DOWSS Time to Clear Environment]
\label{lm:dowss_completion_time}
Let agents in a network executing DOWSS slew their searchlights with
angular speed $\omega$.  Then the time required to clear an environment
with $r$ reflex vertices is no greater than $\frac{2\pi}{\omega}
\frac{1-r^{N}}{1-r}$.
\end{lemma}
\begin{proof}
There are only finitely many ($r$) reflex vertices of $\EE$, and finitely
 many guards ($N$).  Recall each visibility gap encountered during an
 execution of DOWSS produces a semiconvex subregion whose reflex vertices
 necessarily are part of $\partial \EE$. This means the number of visibility
 gaps encountered by any agent when sweeping from $\phi_{\rm start}$ to
 $\phi_{\rm finish}$ (at any level of the recursion tree) can be no greater
 than $r$, i.e., refering to line 8 of PROCESS in
 Tab.~\ref{tab:dowss}, $|G|=m \leq r$.  Since the number of agents
 available to sweep a semiconvex subregion 
 decreases by one for each level of recursion, the maximum depth of the recursion tree is upper bounded by 
 $N-1$.  It is apparent the number of nodes in the recursion tree cannot exceed 
 $1+r+r^2+\dots+r^{N-1} = \frac{1-r^{N}}{1-r}$. 
\end{proof}

\begin{figure}[h]
 \begin{center}
 \resizebox{0.35\linewidth}{!}{\input{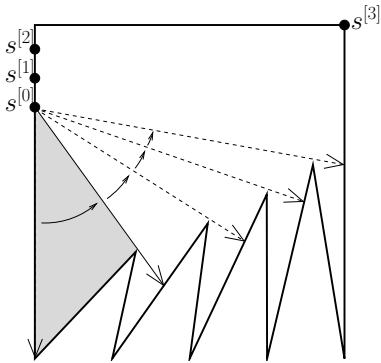}}
 \caption{An example from a class of searchlight instances for which
   malevolent guard choice in OWSS or DOWSS implies time to clear the
   environment is $\mathcal{O}(r^2)$ (and therefore $\mathcal{O}(n^2)$).
   Here $r=4$ reflex vertices are oriented on the bottom so that
   $s^{[r]}=s^{[4]}$ in the upper right corner sees the entire environment.
   $r-1=3$ guards are placed in the upper left and  $s^{[0]}$ is chosen as
   the root.  $s^{[0]}$ clears up to the first reflex vertex (grey) where
   it stops and calls upon $s^{[1]}$ for help.  $s^{[1]} $ then calls upon
   $s^{[2]}$ which likewise calls upon $s^{[3]}$.  This happens every time
   $s^{[0]}$ stops (dashed lines) at the other $r-1$ reflex vertices.  The
   recursion tree of such an execution has $1+r(r-1)$ nodes, thus the
   environment is cleared in $\mathcal{O}(r^2)$ time. }
 \label{fig:anti_owss}
 \end{center}
 \end{figure}

It is not known whether this bound is tight, but at least examples as in
Fig.~\ref{fig:anti_owss} can be constructed where DOWSS and OWSS run in
$\mathcal{O}(r^2)$ ($\Rightarrow \mathcal{O}(n^2)$) time if guards are
chosen malevolently.  A key point is that DOWSS and OWSS do not specify (i)
how to place guards given an environment, or (ii) how to optimally choose
guards at each step given a set of guards.  These are interesting unsolved
problems in their own right which we do not explore in this paper.

 Another performance measure of a distributed algorithm is the size of the
 messages which must be communicated.

\begin{lemma}[DOWSS Message Size]
\label{lm:dowss_message_size}
If the environment has $n$ sides, $r$ reflex vertices, and $N$ agents then the polyline (passed as a message between agents during DOWSS) consists of a list of no more than $\min \{ r+1, N \}$ points in $\real^2$.  Furthermore, since $r \leq n-3$, the list consists of no more than $n-2$ points in $\real^2$.
\end{lemma}
\begin{proof}
For every segment (which is a segment of some searchlight's beam) in such a polyline,  there corresponds a unique reflex vertex of the environment.  The correspondence comes from the fact that at a given time every searchlight supporting a semiconvex subregion has its searchlight aimed at a reflex vertex where it's visibility is occluded.  The uniqueness comes from the fact that if two searchlights support the same semiconvex subregion, say $R$, and are aimed at the same reflex vertex, then only one of the searchlights' beams can actually constitute a portion of $\partial R$ of positive length.  This shows the polyline can consist of no more than $r$ segments and therefore $r+1$ vertices.  Also, in the worst case, the polyline grows by one edge for each level of recursion.  Such polylines start out as a line segment (defined by two points) and the recursion depth cannot exceed $N-1$.  We conclude the maximum number of points defining any polyline is $\min \{ r+1,N \}$.
\end{proof}
 
\begin{table}
\caption{\label{tab:dowss_short} Asynchronous Schedule for Distributed One Way Sweep Strategy (cf Fig. \ref{fig:owss}, \ref{fig:procschedule}, \ref{fig:dowss}, Tab. \ref{tab:dowss}) }
\begin{center}
\noindent{\framebox[.9999\linewidth]{\noindent\parbox{.9999\linewidth-2\fboxsep}{%
\noindent\begin{tabular}{ll}
\textbf{Name:}      &
\parbox[t]{.7\linewidth}{DOWSS}
\\
\textbf{Goal:}      &
\parbox[t]{.7\linewidth}{Agents in the network coordinate their searchlight slewing to clear an environment $\EE$.}
\\
\textbf{Assumes:}            &
\parbox[t]{.7\linewidth}{Agents are stationary and have a completely connected communication topology with no packet loss.  Sweeping is initialized by a root.}
\vspace{0.4em}
\\ \hline
\end{tabular}\\[.5ex]
\\ \\  For time $t>0$, each agent executes the following actions between any two wake up instants according to the schedule in Section~\ref{sec:asynch_model}:
\\ \\ 
{\bf SPEAK}
\begin{algorithmic}[]
        \STATE  Broadcast either
        \STATE \begin{enumerate}
        	             \item a request for help,
	             \item a message to engage a child, or
	             \item a signal of task completion to a parent.
	             \end{enumerate}
\end{algorithmic}

\vspace{.6em}
        
{\bf LISTEN}
\begin{algorithmic}[]
        \STATE Listen for either
        \STATE \begin{enumerate}
                       \item a help request from a potential parent,
                       \item volunteers to help, 
                       \item engagement by parent, or 
                       \item current child reporting completion.
                       \end{enumerate}
\end{algorithmic}

\vspace{0.6em}

{\bf PROCESS}
\begin{algorithmic}[]      
        \STATE \begin{enumerate}
                       \item Use oriented polyline from potential parent with information from sensing to check if 	           
                                able to help, or
	             \item if engaged, compute wayangles, visibility gaps and oriented polylines.
	             \end{enumerate}
\end{algorithmic}
       
\vspace{.6em}
        
{\bf SLEW}
\begin{algorithmic}[]
        \STATE \begin{enumerate}
        		    \item Aim at start angle and switch searchlight on, 
		    \item slew to next angle, or
		    \item slew to finish angle and switch searchlight off.
		    \end{enumerate}
\end{algorithmic}
}}}
\end{center}
\vspace{-3ex}\end{table}

That DOWSS allows flexibility in guard positions (only standing assumptions
required) may be an advantage if agents are immobile.  However, DOWSS only allowing one searchlight slewn at a time is a clear disadvantage when time to clear the environment is to be minimized.  This lead us to
design the algorithm in the next section.

\subsection{Positioning Guards for Parallel Sweeping}
\label{subsec:ptss}

The DOWSS algorithm in the previous section is a distributed
message-passing and local sensing scheme to perform searchlight scheduling
given \emph{a priori} the location of the searchlights. Given an arbitrary
positioning, time to completion of DOWSS can be large;
see Lemma~\ref{lm:dowss_completion_time} and
Figure~\ref{fig:anti_owss}. 

 
The algorithm we design in this section, called the Parallel Tree Sweep
Strategy (PTSS), provides a way of choosing searchlight locations and a
corresponding schedule to achieve faster clearing times.  PTSS works
roughly like this: According to some technical criteria described below,
the environment is partitioned into regions called cells with one agent
located in each cell.  Additionally, the network possesses a distributed
representation of a rooted tree.  By distributed representation we mean that 
every agent knows who its parent and children are.  Using the tree,
agents slew their searchlights in a way that expands the clear region from
the root out to the leaves, thus clearing the entire environment.  Since
agents may operate in parallel, time to clear the environment is linear in
the height of the tree and thus $\mathcal{O}(n)$.
Guaranteed linear time to completion is a clear advantage over DOWSS which can be quadratic or worse (see
Lemma~\ref{lm:dowss_completion_time} and Fig.~\ref{fig:anti_owss}).  Before
describing PTSS more precisely, we need a few definitions.

\begin{definition}
\label{defn:partition_stuff}
\begin{enumerate}
\item A set $\s \subset \real^2$ is \emph{star-shaped} if
    there exists a point $p \in \s$ with the property that all points in
    $\s$ are visible from $p$.  The set of all such points of a given
    star-shaped set $\s$ is called the \emph{kernel} of $\s$ and is denoted
    by $\ker(\s)$. 
\item Given a compact subset $\EE$ of $\real^2$, a \emph{partition} of $\EE$ is
  is a collection of sets $\{\PP^{[0]}, \ldots, \PP^{[N-1]}\}$ such that
  $\union_{i = 0}^{N-1} \PP^{[i]} = \EE$ where $\PP^{[i]}$'s are compact, simply
  connected subsets of $\EE$ with disjoint interiors.  $\{\PP^{[0]}, \ldots,
  \PP^{[N-1]}\}$ will be called \emph{cells} of the partition.
\end{enumerate}
\end{definition}


For our purposes a \emph{gap} (which visibility gap is a special case of)
will refer to any segment $[q,q']$ with $q,q' \in \partial \EE$ and ${]q,q'[}
\in \mathring{\EE}$.  The cells of the partitions we consider will be
separated by gaps.


\begin{definition}[PTSS partition]
\label{defn:ptss_partition}
Given a simple polygonal environment $\EE$, a partition $\{\PP^{[0]}, \ldots,
\PP^{[N-1]}\}$ is a \emph{PTSS partition} if the following conditions are
true:
\begin{enumerate}
\item $\PP^{[i]}$ is a star-shaped cell for all $i \in \{0,\ldots,N-1\}$; 
\item the dual graph\footnote{The dual graph of a partition is the graph
    with cells corresponding to nodes, and there is an edge between nodes
    if the corresponding cells share a curve of nonzero length.} of the
  partition is a tree;
\item a root, say $\PP^{[0]}$, of the dual graph may be chosen so that
  $\ker(\PP^{[0]}) \cap \partial \EE \ne \emptyset$, and for any node other
  than the root, say $\PP^{[k]}$ with parent $\PP^{[j]}$, we have that
  $(\PP^{[j]} \cap \PP^{[k]}) \cap \ker(\PP^{[k]}) \cap \partial \EE \ne
  \emptyset$.
\end{enumerate}
\end{definition}

\begin{definition}
  Given a PTSS partition $\{\PP^{[0]}, \ldots, \PP^{[N-1]}\}$ of $\EE$ and a
  root cell $\PP^{[0]}$ of the partition's dual graph satisfying the
  properties discussed in Definition~\ref{defn:ptss_partition}, the
  corresponding (rooted) \emph{PTSS tree} is defined as follows:
\begin{enumerate}
\item the node set $(p^{[0]},\ldots, p^{[N-1]})$ is such that $p^{[0]} \in
  \ker(\PP^{[0]}) \cap \partial \EE$ and for $k > 1$, $p^{[k]} \in (\PP^{[j]}
  \cap \PP^{[k]}) \cap \ker(\PP^{[k]}) \cap \partial \EE$, where $\PP^{[j]}$
  is the parent of $\PP^{[k]}$ in the dual graph of the partition;
\item there exists an edge $(p^{[j]},p^{[k]})$ if and only if there exists
  an edge $(\PP^{[j]},\PP^{[k]})$ in the dual graph.
\end{enumerate}
\end{definition}

\begin{figure}[h]
\begin{center}
\resizebox{0.46\linewidth}{!}{\includegraphics{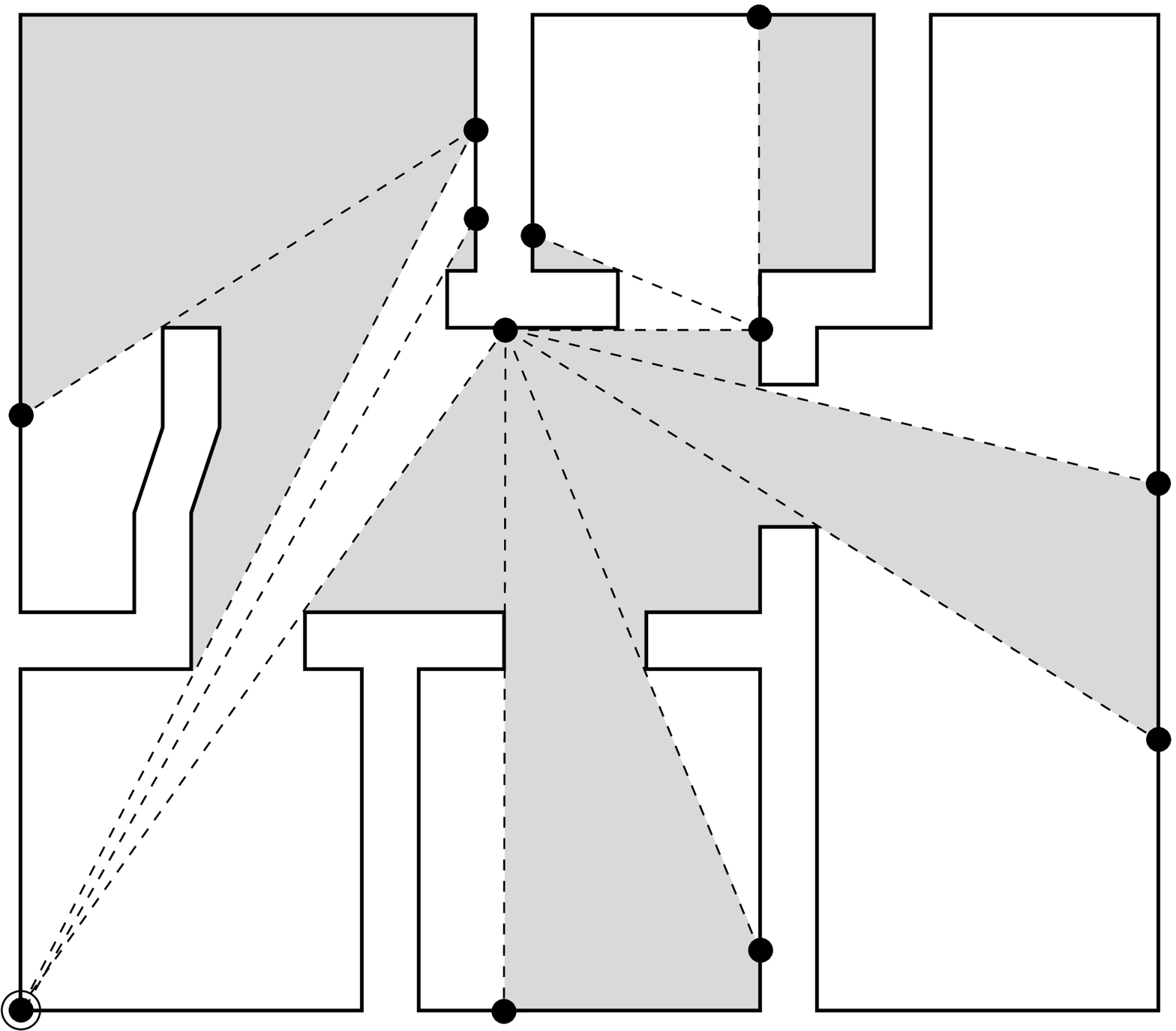} \qquad}
\resizebox{0.455\linewidth}{!}{\includegraphics{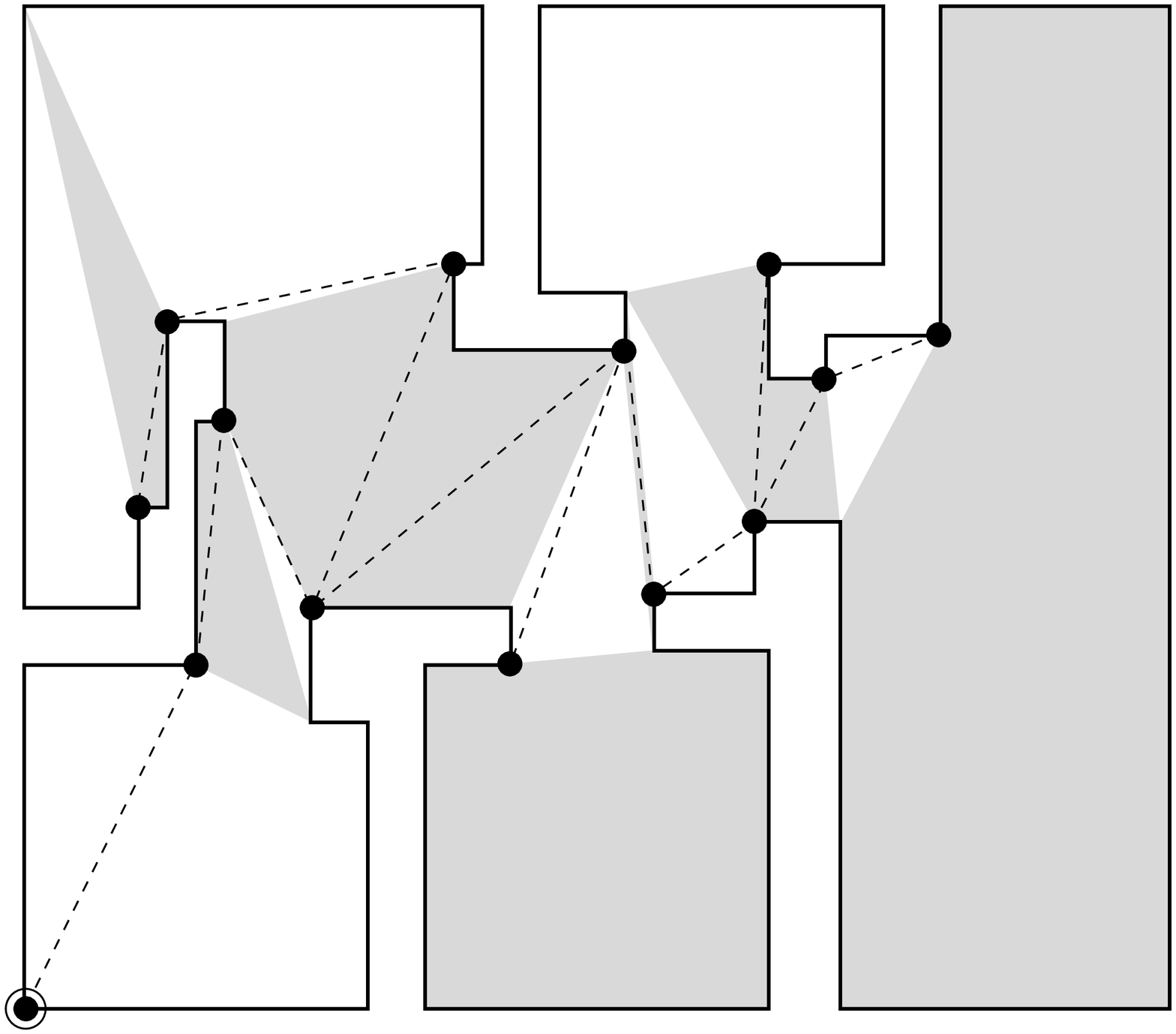} \qquad}
 \caption{Left are agent positions resulting from a Reflex Vertex Straddling (RVS)
  deployment.  Right are agent positions resulting from the deployment described in
  \cite{AG-JC-FB:06r} in which an orthogonal environment is partitioned
  into convex quadrilaterals.  The PTSS partitions are shown by coloring the cells
  alternating grey and white (caution:  grey does not depict clarity here).
  Dotted lines show edges of the PTSS tree where the circled agent is the
  root. }
\label{fig:ptss_partitions}
\end{center}
\end{figure}

We now describe two examples of PTSS partitions seen in
Fig.~\ref{fig:ptss_partitions}.  The left configuration in
Fig.~\ref{fig:ptss_partitions} results from what we call a Reflex Vertex
Straddling (RVS hereinafter) deployment.  RVS deployment begins with all
agents located at the root followed by one agent moving to the furthest end
of each of the root's visibility gaps, thus becoming children of the root.
Likewise, further agents are deployed from each child to take positions on
the furthest end of the children's visibility gaps located across the gaps
dividing the parent from the children.  In this way, the root's cell in the
PTSS partition is just its visibility set, but the cells of all successive
agents consist of the portion of the agents' visibility sets lying across
the gaps dividing their cells from their respective parents' cells.  It is
easy to see that in final positions resulting from an RVS deployment,
agents see the entire environment.
\begin{lemma}
\label{lm:rvs_guards}
RVS deployment requires, in general, no more than $r+1 \leq n-2$ agents to see the entire environment from their final positions.  In an orthogonal environment, no more than $\frac{n}{2}-2$ agents are required.
\end{lemma}
\begin{proof}
Follows from the fact that in addition to the root, no more than one agent will be placed for each reflex vertex (only reflex vertices occlude visibility).
\end{proof}

See Fig.~\ref{fig:ptss_sim} for simulation results of PTSS executed by
agents in an RVS configuration.  The right configuration in
Fig.~\ref{fig:ptss_partitions} results from the deployment described in
\cite{AG-JC-FB:06r} in which an orthogonal environment is partitioned into
convex quadrilaterals.
\begin{lemma}
\label{lm:ag_guards}
The deployment described in \cite{AG-JC-FB:06r} requires no more than $\frac{n}{2}-2$ agents to see the entire (orthogonal) environment from their final positions.  
\end{lemma}
\begin{proof}
See \cite{AG-JC-FB:06r}.
\end{proof}

Both of the PTSS configurations in these examples 
may be generated via distributed deployment algorithms in which agents
perform a depth-first, breadth-first, or randomized search on the PTSS tree
constructed on-line.  Please refer to \cite{AG-JC-FB:05z} and
\cite{AG-JC-FB:06r} for a detailed description of these algorithms.

We now turn our attention to the pseudocode in Tab.~\ref{tab:ptss_short}  (A more detailed pseudocode, which we refer to in the proofs, can be found in the appendix, Tab.~\ref{tab:ptss}) and describe PTSS more precisely.  Suppose some
agents are positioned in an environment according to a PTSS partition and
tree with agent $1$ as the root.  PTSS begins by agent $1$ pointing its
searchlight along a wall in the direction $\phi_{\rm start}$ and then
slewing away from the wall toward $\phi_{\rm finish}$, pausing whenever it
encounters the first side of a gap, say $\phi_j$, where $j$ is odd.  Paused
at $\phi_j$, agent $1$ sends a message to its child at that gap, say agent
$2$, so that agent $2$ knows it should aim its searchlight across the gap.
Once agent $2$ has its searchlight safely aimed across the gap, it sends a
message to agent $1$ so that agent $1$ knows it may continue slewing over
the whole gap.  When agent $1$ has reached the other side of the gap at
$\phi_{j+1}$, agent $1$ sends a message to agent $2$ and both agents
continue clearing the rest of their cells concurrently, stopping at gaps
and coordinating with children as necessary.  In this way, the clear region
expands from the root to the leaves at which time the entire environment
has been cleared.  We arrive at the following lemmas and correctness result.

\begin{figure}[h]
\begin{center}
\resizebox{0.3\linewidth}{!}{ \input{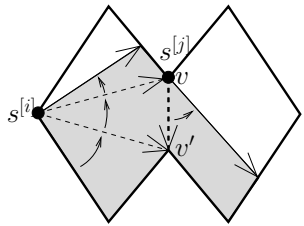} }
\caption{Expanding a clear region (grey) across a gap (thick dashed segment
  [v, v']) from cell $\PP^{[i]}$ to cell $\PP^{[j]}$ may always be
  accomplished by the child ($s^{[j]}$) aiming across the gap and waiting
  for the parent ($s^{[i]}$) to slew over the gap.  Both agents may then
  continue clearing the remainder of their respective cells. }
\label{fig:gap_pass}
\end{center}
\end{figure}

\begin{lemma}[Expanding a Clear Region Across a Gap]
\label{lm:gap_pass}
Suppose an environment is endowed with a PTSS partition and tree, and that
agent $i$ is a parent of agent $j$ (see Fig.~\ref{fig:gap_pass}).  Then a
clear region may always be expanded across the gap from $\PP^{[i]}$ to
$\PP^{[j]}$ by $s^{[j]}$ first aiming across the gap and waiting for
$s^{[i]}$ to slew over the gap.  Both agents may then continue clearing the
remainder of their respective cells concurrently.
\end{lemma}
\begin{proof}
This obviously hold for the scenario in Fig.~\ref{fig:gap_pass}.  Using the definition of PTSS partition, it is clear any general PTSS parent-child relationship is reducible to the case in Fig.~\ref{fig:gap_pass}.
\end{proof}

\begin{theorem}[Correctness of PTSS]
\label{thm:ptss_correctness}
Given a simple polygonal environment $\EE$ and agent positions
$P=(p^{[0]},\ldots, p^{[N-1]})$, let the following conditions hold:
\begin{enumerate}
\item the standing assumptions are satisfied;
\item all agents $i \in \{0,\ldots,N-1\}$ are positioned in a PTSS partition and rooted tree with agent $1$ as the root;
\item the agents operate under PTSS. 
\end{enumerate}
Then $\EE$ is cleared in finite time.
\end{theorem}
\begin{proof}
Follows immediately from Lemma~\ref{lm:gap_pass}.
\end{proof}

Since multiple branches of the PTSS tree may be cleared concurrently, and using Lemmas \ref{lm:rvs_guards} and \ref{lm:ag_guards}, we have the next lemma (assuming processing and communication time are negligible, cf. Lemma~\ref{lm:dowss_completion_time}).

\begin{lemma}[PTSS Time to Clear Environment]
\label{lm:ptss_completion_time}
Let the agents in a network executing PTSS slew their searchlights with
angular speed $\omega$.  Then time required to clear an environment is
\begin{enumerate}
\item  linear in the height of the PTSS tree; 
\item no greater than $\frac{2 \pi}{\omega} (r+1) \leq \frac{2 \pi}{\omega}
  (n-2)$ if agents are in final positions according to an RVS deployment;
\item no greater than $\frac{\pi}{\omega} (n-2)$ if agents are in final
  positions in an orthogonal polygon according to an RVS deployment or the
  deployment described in \cite{AG-JC-FB:06r}.
\end{enumerate}
\end{lemma}
\begin{proof}
With communication time neglibile, each child will wait for it's parent a maximum time of  $\frac{2 \pi}{\omega}$.  It now suffices to observe that the maximum length of any parent-child sequence is just the height of the PTSS tree.
\end{proof}

Looking at the SPEAK section of Tab.~\ref{tab:ptss}, it is easy to see
that message size is constant (cf. Lemma~\ref{lm:dowss_message_size}).
\begin{lemma}[PTSS Message Size]
\label{lm:ptss_message_size}
Messages passed between agents executing PTSS have constant size.
\end{lemma}

\begin{table}
\caption{\label{tab:ptss_short} Asynchronous Schedule for Parallel Tree Sweep Strategy (cf Fig. \ref{fig:procschedule}, \ref{fig:gap_pass}, \ref{fig:ptss_partitions}, Tab. \ref{tab:ptss}) }
\begin{center}
\noindent{\framebox[.9999\linewidth]{\noindent\parbox{.9999\linewidth-2\fboxsep}{%
\noindent\begin{tabular}{ll}
\textbf{Name:}      &
\parbox[t]{.7\linewidth}{PTSS}
\\
\textbf{Goal:}      &
\parbox[t]{.7\linewidth}{Agents in the network coordinate their searchlight slewing to clear an environment $\EE$.}
\\
\textbf{Assumes:}       &
\parbox[t]{.7\linewidth}{Agents are statically positioned as nodes in a PTSS partition and tree, and each knows a priori the gaps of its cell and UIDs of the corresponding children and parent.  Sweeping is initialized by the root.}
\vspace{0.4em}
\\ \hline
\end{tabular}\\[.5ex]
\\ \\  For time $t>0$, each agent executes the following actions between any two wake up instants according to the schedule in Section~\ref{sec:asynch_model}:
\\ \\ 
{\bf SPEAK}
\begin{algorithmic}[]
        \STATE Broadcast either
                      \begin{enumerate}
                      \item a command for a child to aim across a gap, 
                      \item a confirmation to a parent when aimed across gap, or
                      \item when finished slewing over a gap, a signal of completion to the child.
                      \end{enumerate}
\end{algorithmic}

\vspace{.6em}
        
{\bf LISTEN}
\begin{algorithmic}[]
        \STATE Listen for either 
                      \begin{enumerate}
                      \item instruction from a parent to aim across a gap,
                      \item confirmation from a child aimed across a gap, or
                      \item confirmation that parent has passed the gap.
                      \end{enumerate}
\end{algorithmic}

\vspace{0.6em}

{\bf PROCESS}
\begin{algorithmic}[]
        \STATE When first engaged, compute wayangles where coordination with children will be necessary.
\end{algorithmic}
       
\vspace{.6em}
        
{\bf SLEW}
\begin{algorithmic}[]
        \STATE \begin{enumerate}
                      \item Aim at start angle and switch searchlight on, 
                      \item slew to next wayangle, or
                      \item slew to finish angle and switch searchlight off.
                      \end{enumerate}
\end{algorithmic}
}}}
\end{center}
\vspace{-3ex}\end{table}

Requiring guards to be situated in a PTSS tree may be more restrictive than
the mere standing assumptions required by DOWSS, but the time savings using
PTSS over DOWSS can be considerable.  Though we have given two examples of
how to construct a PTSS tree, it is not clear how to construct one which clears an environment in minimum time among all possible PTSS trees.  It is also not clear how to optimally choose the root of the tree (point of deployment).  However, if information about an environment layout is known a priori and one may choose the root location, then an exhaustive strategy may be adopted whereby all possible root choices are compared.


\section{Conclusions}
\label{sec:conclusion}

In this paper we have provided two solutions to the distributed searchlight
scheduling problem.  DOWSS requires guards satisfying the standing
assumptions, has message size $\mathcal{O}(n)$, and sometimes takes time
$\mathcal{O}(r^2)$ to clear an environment.  PTSS requires agents are
positioned according to a PTSS tree, has constant message size, and takes
time linear in the height of the PTSS tree to clear the environment.  We
have given two procedures for constructing PTSS trees, one requiring no
more than $r \leq n-3$ guards for a general polygonal environment, and two
requiring no more than $\frac{n-2}{2}$ guards for an orthogonal
environment.  Guards slew through a total angle no greater than $2 \pi$, so
the upper bounds on the time for PTSS to clear an environment with these
partitions are $\frac{2 \pi}{\omega} r \leq \frac{2 \pi}{\omega} (n-3)$ and
$\frac{\pi}{\omega}(n-2)$, respectively.  Because PTSS allows searchlights
to slew concurrently, it generally clears an environment much faster than
DOWSS.  However, the comparison is not completely fair since DOWSS does not
specify how to choose guards but PTSS does.

To extend DOWSS and PTSS for environments with holes, one simple solution
is to add one guard per hole, where a simply connected environment is
simulated by the extra guards using their beams to connect the holes to the
outer boundary.  Another straightforward extension for PTSS would be to
combine it directly with a distributed deployment algorithm such as those
in \cite{AG-JC-FB:05z} and \cite{AG-JC-FB:06r}, so that deployment and
searchlight slewing happen concurrently.  This suggests an interesting
problem we hope to explore in the future, namely minimizing the time to
perform a coordinated search given a limited number of mobile guards.
Other considerations for the future include loosening the requirements in
the definition of the PTSS partition, and incorporating in our model sensor
constraints such as limited depth of field and beam incidence.


{\small


\begin{thebibliography}{10}

\bibitem{DPB-JNT:97}
{\sc D.~P. Bertsekas and J.~N. Tsitsiklis}, {\em Parallel and Distributed
  Computation: Numerical Methods}, Athena Scientific, Belmont, MA, 1997.

\bibitem{AG-JC-FB:05z}
{\sc A.~Ganguli, J.~Cort{\'e}s, and F.~Bullo}, {\em Distributed deployment of
  asynchronous guards in art galleries}, in {A}merican {C}ontrol {C}onference,
  Minneapolis, MN, June 2006, pp.~1416--1421.

\bibitem{AG-JC-FB:06r}
\leavevmode\vrule height 2pt depth -1.6pt width 23pt, {\em Visibility-based
  multi-agent deployment in orthogonal environments}, in {A}merican {C}ontrol
  {C}onference, New York, July 2007.
\newblock To appear.

\bibitem{BPG-ST-GG:06}
{\sc B.~P. Gerkey, S.~Thrun, and G.~Gordon}, {\em Visibility-based
  pursuit-evasion with limited field of view}, International Journal of
  Robotics Research, 25 (2006), pp.~299--315.

\bibitem{JHL-SMP-KYC:02}
{\sc J.~H. Lee, S.~M. Park, and K.~Y. Chwa}, {\em Simple algorithms for
  searching a polygon with flashlights}, Information Processing Letters, 81
  (2002), pp.~265--270.

\bibitem{JOR:87}
{\sc J.~O'Rourke}, {\em Art Gallery Theorems and Algorithms}, Oxford University
  Press, Oxford, UK, 1987.

\bibitem{TCS:92}
{\sc T.~C. Shermer}, {\em Recent results in art galleries}, IEEE Proceedings,
  80 (1992), pp.~1384--1399.

\bibitem{BS-GS-SML:00}
{\sc B.~Simov, G.~Slutzki, and S.~M. LaValle}, {\em Pursuit-evasion using beam
  detection}, in {IEEE} Int. Conf. on Robotics and Automation, 2000.

\bibitem{KS-IS-MY:90}
{\sc K.~Sugihara, I.~Suzuki, and M.~Yamashita}, {\em The searchlight scheduling
  problem}, SIAM Journal on Computing, 19 (1990), pp.~1024--1040.

\bibitem{JU:00}
{\sc J.~Urrutia}, {\em Art gallery and illumination problems}, in Handbook of
  Computational Geometry, J.~R. Sack and J.~Urrutia, eds., North-Holland,
  Amsterdam, the Netherlands, 2000, pp.~973--1027.

\bibitem{MY-IS-TK:04}
{\sc M.~Yamashita, I.~Suzuki, and T.~Kameda}, {\em Searching a polygonal region
  by a group of stationary k-searchers}, Information Processing Letters, 92
  (2004), pp.~1--8.

\bibitem{MY-HU-IS-TK:01}
{\sc M.~Yamashita, H.~Umemoto, I.~Suzuki, and T.~Kameda}, {\em Searching for
  mobile intruders in a polygonal region by a group of mobile searchers},
  Algorithmica, 31 (2001), pp.~208--236.

\end{thebibliography}
}

\clearpage 
\section{Appendix:  Extended versions of Tab.~\ref{tab:dowss_short} and \ref{tab:ptss_short} referred to in proofs}

\begin{table}
\caption{\label{tab:dowss} Asynchronous Schedule for Distributed One Way Sweep Strategy (cf Fig. \ref{fig:owss}, \ref{fig:procschedule}, \ref{fig:dowss}, Tab. \ref{tab:dowss_short}) }
\begin{center}
\noindent{\framebox[.9999\linewidth]{\noindent\parbox{.9999\linewidth-2\fboxsep}{%
\noindent\begin{tabular}{ll}
\textbf{Name:}      &
\parbox[t]{.8\linewidth}{DOWSS}
\\
\textbf{Goal:}      &
\parbox[t]{.8\linewidth}{Agents in the network coordinate their searchlight slewing to clear an environment $Q$.}
\\
\textbf{Assumes:}            &
\parbox[t]{.8\linewidth}{Agents are stationary and have a completely connected communication topology with no packet loss.  Sweeping is initialized by a root who's uid is 1.}
\vspace{0.4em}
\\ \hline
\end{tabular}\\[.5ex]

The root initially has ${\rm state}=4$ and all other agents begin with ${\rm state}=1$, where possible states are $1, 2, 3,\ldots,12$.  Aditionally, each agent $i \in \{0,\ldots, N-1\}$, possesses local variables ${\rm parent}$, ${\rm child}$, $\psi_{\rm parent}$, $\psi_{\rm temp}$, $j$, $G$, $\Phi$, $\phi_{\rm start}$, $\phi_{\rm finish}$, $\Psi$, and $u$, all initially empty.    As needed to clarify ownership, a superscript with square brackets indicates the UID of the agent to whom a variable belongs.
\\ \\  For time $t>0$, each agent $i$ executes the following between any two wake up instants according to the schedule in Section~\ref{sec:asynch_model}:
\\
\begin{center} 
\begin{tabular}{l  l}
\parbox[t]{.46\linewidth}{
{\tiny
{\bf SPEAK}
\begin{algorithmic}[1]
\WHILE{${\rm state} = 7$ or ${\rm state} = 8$}
        \STATE \COMMENT{request help}
        \STATE $\BROADCAST(i, \psi_j(t), {\tt help})$
        \STATE ${\rm state} \leftarrow 8$
\ENDWHILE
\WHILE{${\rm state}=2$}
        \STATE \COMMENT{volunteer to help}
        \STATE $\BROADCAST(i, {\rm parent}(t), \psi_{\rm temp}(t), {\tt volunteer})$
        \STATE ${\rm state} \leftarrow 3$
\ENDWHILE
\WHILE{${\rm state} = 9$}
        \STATE \COMMENT{engage a child}
        \STATE $\BROADCAST(i, {\rm child}(t), \psi_j(t), {\tt selected})$
        \STATE ${\rm state} \leftarrow 10$
\ENDWHILE
\WHILE{${\rm state}=12$}
        \STATE \COMMENT{report to parent when complete}
        \STATE $\BROADCAST(i, {\rm parent}, \psi_{\rm parent}(t), {\tt complete})$
        \STATE ${\rm state} \leftarrow 1$
\ENDWHILE
\end{algorithmic}

\vspace{.6em}
        
{\bf LISTEN}
\begin{algorithmic}[1]
\WHILE{${\rm state} = 1$ or ${\rm state} = 3$}
        \STATE \COMMENT{listen for help request}
        \IF{$\RECEIVE(i', \psi^{[i']}_{j'}(t-\tau), {\tt help})$, where $0 \le \tau \le \delta$}
        \STATE $\psi_{\rm temp} \leftarrow \psi^{[i']}_{j'}(t-\tau)$; ${\rm state} \leftarrow 3$
        \ENDIF
\ENDWHILE
\WHILE{${\rm state} = 8$}
        \STATE \COMMENT{listen for volunteers}
        \IF{$\RECEIVE(i', {\rm parent}^{[i']}(t-\tau), \psi^{[i']}_{\rm temp}(t-\tau), {\tt volunteer})$, where $0 \le \tau \le \delta$, ${\rm parent}^{[i']}(t-\tau) = i$, and  $\psi^{[i']}_{\rm temp}(t-\tau)=\psi_{j}$}
        \STATE ${\rm child} \leftarrow i' ; {\rm state} \leftarrow 9$
        \ENDIF
\ENDWHILE
\WHILE{${\rm state} = 3$} 
        \STATE \COMMENT{listen for engagement by parent}
        \IF{$\RECEIVE(i', {\rm child}^{[i']}, \psi^{[i']}_{j'}(t-\tau), {\tt selected})$, where $0 \le \tau \le \delta$, where ${\rm child}^{i'}=i$}
        \STATE ${\rm parent} \leftarrow i' ; \psi_{\rm parent} \leftarrow \psi^{[i']}_{j'}(t-\tau);{\rm state} \leftarrow 4$
        \ELSIF{$\RECEIVE(i', {\rm child}^{[i']}, \psi^{[i']}_{j'}(t-\tau), {\tt selected})$, where $0 \le \tau \le \delta$, where ${\rm child}^{[i']} \neq i$}
        \STATE ${\rm state} \leftarrow 1$
        \ENDIF
\ENDWHILE
\WHILE{${\rm state} = 10$}
        \STATE \COMMENT{listen for child to report completion}
        \IF{ $\RECEIVE(i', {\rm parent}^{[i']}, \psi^{[i']}_{\rm parent}(t-\tau), {\tt complete})$, where $0 \le \tau \le \delta$ }
                \IF{$j<m$} 
                        \STATE $j \leftarrow j+1; {\rm state} \leftarrow 6$ 
                \ELSIF{$j=m$} 
                        \STATE ${\rm state} \leftarrow 11$
                \ENDIF
        \ENDIF
\ENDWHILE
\end{algorithmic}

}} 
&
\parbox[t]{.46\linewidth}{
{\tiny
{\bf PROCESS}
\begin{algorithmic}[1]
\WHILE{${\rm state}=3$}        
        \STATE \COMMENT{use $\psi_{\rm temp}$ and $\VV(p^{[i]})$ to check if able to help}
        \IF{able to see across oriented polyline $\psi_{\rm temp}$ into semiconvex subregion and not located in interior of that subregion}
                \STATE ${\rm state} \leftarrow 2$
        \ENDIF
\ENDWHILE
\WHILE{${\rm state}=4$} 
        \STATE \COMMENT{when first engaged, perform geometric computations;  note visibility gaps are listed ccw and radially outwards}
        \STATE Compute $\phi_{\rm start}$ and $\phi_{\rm finish}$ \COMMENT{start and finish angles}
        \STATE Compute $G \leftarrow ( g_1, \ldots, g_m )$ \COMMENT{visibility gaps}
        \STATE Compute $\Phi \leftarrow (\phi_1, \ldots, \phi_m)$ \COMMENT{resp. angles of visibility gaps}
        \STATE Compute $\Psi \leftarrow ( \psi_1, \ldots, \psi_m )$ \COMMENT{polyline for each visibility gap}
        \STATE $j \leftarrow 1$ \COMMENT{initialize slewing counter}
        \STATE ${\rm state} \leftarrow 5$
\ENDWHILE
\end{algorithmic}
       
\vspace{.5em}
        
{\bf SLEW}
\begin{algorithmic}[1]
\WHILE{${\rm state}=5$} 
        \STATE \COMMENT{aim at start angle and switch searchlight on}
        \STATE $\theta^{[i]} \leftarrow \phi_{\rm start}$
        \STATE ${\rm state} \leftarrow 6$
\ENDWHILE

\WHILE{${\rm state}=6$} 
        \STATE \COMMENT{slew to next angle}
        \WHILE{$\theta^{[i]}<\phi_j$}
                        \STATE $u \leftarrow \frac{\min\{\smax, ||\phi_j-\theta^{[i]}||\}}{||\phi_j-\theta^{[i]}||}(\phi_j-\theta^{[i]})$
                \STATE $\theta^{[i]} \leftarrow \theta^{[i]}+u$
                \ENDWHILE
        \STATE ${\rm state} \leftarrow 7$
\ENDWHILE

\WHILE{${\rm state}=11$} 
        \STATE \COMMENT{slew to finish angle and switch searchlight off}
        \WHILE{$\theta^{[i]}<\phi_{\rm finish}$}
                        \STATE $u \leftarrow \frac{\min\{\smax, ||\phi_{\rm finish}-\theta^{[i]}||\}}{||\phi_{\rm finish}-\theta^{[i]}||}(\phi_{\rm finish}-\theta^{[i]})$
                \STATE $\theta^{[i]} \leftarrow \theta^{[i]}+u$
        \ENDWHILE
        \STATE ${\rm state}\leftarrow 12$
\ENDWHILE
\end{algorithmic}

}} 
\end{tabular} 
\end{center}

\vspace{0.5em}

}}}
\end{center}
\vspace{-3ex}\end{table}
  

\begin{table}
\caption{\label{tab:ptss} Asynchronous Schedule for Parallel Tree Sweep Strategy (cf Fig. \ref{fig:procschedule}, \ref{fig:gap_pass}, \ref{fig:ptss_partitions}, Tab. \ref{tab:ptss_short}) }
\begin{center}
\noindent{\framebox[.9999\linewidth]{\noindent\parbox{.9999\linewidth-2\fboxsep}{%
\noindent\begin{tabular}{ll}
\textbf{Name:}      &
\parbox[t]{.8\linewidth}{PTSS}
\\
\textbf{Goal:}      &
\parbox[t]{.8\linewidth}{Agents in the network coordinate their searchlight slewing to clear an environment $Q$.}
\\
\textbf{Assumes:}       &
\parbox[t]{.8\linewidth}{Agents are statically positioned as nodes in a PTSS partition and tree, and each knows a priori the gaps of its cell and UIDs of the corresponding children and parent.  Sweeping is initialized by a root who's UID is 1.  Agents need only communicate with their parents and children.}
\vspace{0.4em}
\\ \hline
\end{tabular}\\[.5ex]

The root initially has ${\rm state}= 2$ and all other agents begin with ${\rm state}= 1$, where possible states are $1, 2,\ldots,10$.  Additionally each agent $i \in \{0,\ldots, N-1\}$, possesses local variables ${\rm parent}$, $\Phi$, $\phi_{\rm start}$, $\phi_{\rm finish}$, $C$, $j$, and $u$, all initially empty.  As needed to clarify ownership, a superscript with square brackets indicates the UID of the agent to whom a variable belongs.
\\ \\  For time $t>0$, each agent $i$ executes the following between any two wake up instants according to the schedule in Section~\ref{sec:asynch_model}:
\\
\begin{center} 
\begin{tabular}{l  l}
\parbox[t]{.46\linewidth}{
{\tiny
{\bf SPEAK}
\begin{algorithmic}[1]
\WHILE{${\rm state}= 7$}
        \STATE \COMMENT{tell child to aim across gap}
        \STATE $\BROADCAST({\rm child}_j, {\tt aim\_across\_gap})$
        \STATE ${\rm state} \leftarrow 8$
\ENDWHILE
\WHILE{${\rm state}= 4$}
        \STATE \COMMENT{tell parent when aimed across gap}
        \STATE $\BROADCAST(i, {\tt aimed\_across\_gap})$
        \STATE ${\rm state} \leftarrow 5$
\ENDWHILE
\WHILE{${\rm state} = 9$}
        \STATE \COMMENT{tell child when finished slewing over gap}
        \STATE $\BROADCAST({\rm child}_j, {\tt gap\_passed})$
        \IF{$j<m$} 
                        \STATE $j \leftarrow j+1; {\rm state} \leftarrow 6$ 
        \ELSIF{$j=m$} 
                \STATE ${\rm state} \leftarrow 10$
        \ENDIF
\ENDWHILE
\end{algorithmic}

\vspace{.6em}
        
{\bf LISTEN}
\begin{algorithmic}[1]
\WHILE{${\rm state}= 1$}
        \STATE \COMMENT{listen for instruction from parent to aim across gap}
        \IF{$\RECEIVE({\rm child}^{[i']}, {\tt aim\_across\_gap})$ and $i={\rm child}^{[i']}$}
                        \STATE ${\rm state} \leftarrow 2$
        \ENDIF
\ENDWHILE
\WHILE{${\rm state}= 8$}
        \STATE \COMMENT{listen for confirmation from child aimed across gap}
        \IF{$\RECEIVE(i', {\tt aimed\_across\_gap})$ and $i'={\rm child}_j$}
                \STATE $j \leftarrow j+1; {\rm state} \leftarrow 6$ 
       \ENDIF
\ENDWHILE
\WHILE{${\rm state} = 5$} 
        \STATE \COMMENT{listen for confirmation that parent has passed the gap}
        \IF{$\RECEIVE({\rm child}^{[i']}, {\tt gap\_passed})$ and $i={\rm child}^{[i']}$}
        \STATE ${\rm state} \leftarrow 6$
        \ENDIF
\ENDWHILE
\end{algorithmic}
}} 
&
\parbox[t]{.46\linewidth}{
{\tiny 
{\bf PROCESS}
\begin{algorithmic}[1]
\WHILE{${\rm state}= 2$} 
        \STATE \COMMENT{when first engaged, perform geometric computations}
        \STATE Compute $\phi_{\rm start}$ and $\phi_{\rm finish}$ \COMMENT{start and finish angles}
        \STATE Compute $\Phi \leftarrow (\phi_1, \ldots, \phi_m)$ \COMMENT{ordered gap endpoint angles}
        \STATE Compute $C \leftarrow ({\rm child}_1,\ldots, {\rm child}_m)$ \COMMENT{resp. child UIDs}
        \STATE $j \leftarrow 1$;  \COMMENT{initialize slewing counter}
        \STATE ${\rm state} \leftarrow 3$
\ENDWHILE
 \end{algorithmic}
       
\vspace{.5em}
        
{\bf SLEW}
\begin{algorithmic}[1]
\WHILE{${\rm state}=3$} 
        \STATE \COMMENT{aim at start angle and switch searchlight on}
        \STATE $\theta^{[i]} \leftarrow \phi_{\rm start}$
        \STATE ${\rm state} \leftarrow 4$
\ENDWHILE

\WHILE{${\rm state}= 6$} 
        \STATE \COMMENT{slew to next angle}
        \WHILE{$\theta^{[i]}<\phi_j$}
                        \STATE $u \leftarrow \frac{\min\{\smax, ||\phi_j-\theta^{[i]}||\}}{||\phi_j-\theta^{[i]}||}(\phi_j-\theta^{[i]})$
                \STATE $\theta^{[i]} \leftarrow \theta^{[i]}+u$
                \ENDWHILE
        \IF{$j$ is odd}         
        \STATE ${\rm state} \leftarrow 7$
        \ELSIF{$j$ is even}
        \STATE ${\rm state} \leftarrow 9$
        \ENDIF
\ENDWHILE

\WHILE{${\rm state}= 10$} 
        \STATE \COMMENT{slew to finish angle and switch searchlight off}
        \WHILE{$\theta^{[i]}<\phi_{\rm finish}$}
                        \STATE $u \leftarrow \frac{\min\{\smax, ||\phi_{\rm finish}-\theta^{[i]}||\}}{||\phi_{\rm finish}-\theta^{[i]}||}(\phi_{\rm finish}-\theta^{[i]})$
                \STATE $\theta^{[i]} \leftarrow \theta^{[i]}+u$
        \ENDWHILE
        \STATE ${\rm state} \leftarrow 1$
\ENDWHILE
\end{algorithmic}
}} 
\end{tabular} 
\end{center}

\vspace{0.5em}

}}}
\end{center}
\vspace{-3ex}\end{table}
  

\end{document}